\documentclass[conference]{IEEEtran}
\IEEEoverridecommandlockouts
\usepackage{cite}
\usepackage{amsmath}
\usepackage{amssymb,amsfonts}
\usepackage{algorithmic}
\usepackage{graphicx}
\usepackage{textcomp}
\usepackage{xcolor}
\usepackage{booktabs}
\usepackage[colorlinks,
linkcolor=red,       
anchorcolor=red,  
citecolor=blue,        
]{hyperref}

\usepackage{subcaption}
\usepackage{cleveref}

\usepackage{amsthm}
\usepackage[inline]{enumitem}
\theoremstyle{remark}

\newcommand{\DC}{\mathrm{DC}}
\newcommand{\IT}{\mathrm{IT}}
\newcommand{\shed}{\mathrm{shed}}
\newcommand{\idle}{\mathrm{idle}}
\newcommand{\full}{\mathrm{full}}
\newcommand{\PUE}{\mathrm{PUE}}
\newcommand{\MSE}{\mathrm{MSE}}
\newcommand{\cost}{\mathrm{cost}}
\newcommand{\hist}{\mathrm{hist}}

\def\BibTeX{{\rm B\kern-.05em{\sc i\kern-.025em b}\kern-.08em
    T\kern-.1667em\lower.7ex\hbox{E}\kern-.125emX}}
\begin{document}

\title{A Predict-then-Schedule framework for Power Distribution Networks with AI Data Centers\\
\thanks{Siqi Yan and Jiebao Zhang contributed equally to this work. Corresponding authors: Ye Shi and Xi Yao.
}
}



\author{Siqi Yan, Jiebao Zhang, Xi Yao, Juan Huang and Ye Shi
\thanks{Siqi Yan, Jiebao Zhang and Ye Shi are with the School of Information Science and Technology, ShanghaiTech University, Shanghai 201210, China. (email:{\tt yansq2024, zhangjb2023, shiye@shanghaitech.edu.cn})}
\thanks{Xi Yao and Juan Huang are with China Mobile Shanghai ICT Co., Ltd. (email: {\tt yaoxi, huangjuan@cmsr.chinamobile.com})} 
\thanks{This work was supported by National Natural Science Foundation of China (62303319), ShanghaiTech AI4S Initiative SHTAI4S202404, HPC Platform of ShanghaiTech University, and MoE Key Laboratory of Intelligent Perception and Human-Machine Collaboration (ShanghaiTech University).}
}

\maketitle

\begin{abstract}
The surge of GPU-intensive workloads in artificial intelligence (AI) data centers drives massive energy demands, leading to soaring costs and significant stress on local power distribution networks.
Coordinating delay-tolerant workload scheduling with power grid conditions via precise workload prediction can mitigate these issues.
However, a critical gap remains in conventional approaches, i.e., minimizing prediction error does not necessarily lead to minimized downstream operational loss. Hence, this paper proposes an end-to-end Predict-Then-Schedule (PTS) framework that integrates upstream workload prediction with downstream scheduling optimization.
By leveraging differentiable convex optimization, the PTS framework maps input features directly to optimal scheduling and enables gradient-based training. 
Furthermore, to respect the data center's capacity, a workload over-shifted loss combining electricity cost with a penalty for load-shedding is introduced to evaluate scheduling quality.
Experiments demonstrate that the proposed framework significantly reduces operational cost and enhances system security compared to the conventional two-stage baseline.
\end{abstract} 

\begin{IEEEkeywords}
AI Data Center, Power Distribution Network, Workload Prediction, Predict-then-Schedule(PTS)
\end{IEEEkeywords}

\section{Introduction}
The explosive growth of GPU-intensive workloads, such
 as deep learning jobs, has driven the expansion of artificial intelligence (AI) data centers~\cite{masanet2020recalibrating, strubell2019energy}.
{
The large and fluctuating power demand of AI data centers poses significant challenges to distribution grids, leading to higher electricity costs~\cite{sun2022data, munir2021risk} and increasing the risk of grid instability due to capacity constraints~\cite{chen2025electricitydemandgridimpacts}.
}
A promising strategy has emerged to coordinate delay-tolerant workloads with power grid conditions, utilizing their inherent spatio-temporal flexibility to reduce operational costs and enhance stability~\cite{li2016toward, safari2024artificial}.
However, this coordination strategy is critically dependent on precise workload forecasts~\cite{chen2019towards}.
Inaccurate predictions can lead to severe consequences, including high operational costs, energy waste, and distribution power grid instability~\cite{dimd2022review, schmitt2021cost}.
\par
The predict-then-optimize (PTO) framework~\cite{elmachtoub2022smart, donti2017task} aims to avoid the consequences arising from prediction bias.
Unlike the conventional two-stage method, which treats prediction and optimization as two separate, sequential procedures, the PTO framework integrates the downstream optimization problem into the training loop of the upstream predictor \cite{zhangDecisionFocusedLearningPower2025}.
This allows the predictor to produce forecasts that are specifically tailored to high-quality downstream decisions \cite{wilder2019melding}.
\par
In this paper, we propose an end-to-end, predict-then-schedule (PTS) framework that incorporates the downstream workload scheduling optimization directly into the upstream workload prediction training pipeline.
We compute the gradients of the optimal transfer matrix with respect to the upstream predictor's parameters, inspired from differentiable convex programming.
Furthermore, we introduce a workload over-shifted loss, which combines power cost and a penalty term of load-shedding variables. 
This loss evaluates the quality of downstream decisions, thereby guiding prediction results to satisfy physical constraints of power distribution networks and AI data centers.

Our contributions are summarized as follows:
\begin{enumerate*}
    \item To the best of our knowledge, we are the first to integrate downstream workload scheduling optimization into the training of workload prediction models in an end-to-end differentiable manner. Unlike the conventional two-stage method, the proposed framework leverages the sensitivity of scheduling decisions on upstream predictors, enabling workload forecasts that are aware of scheduling quality.
    \item We propose a novel workload over-shifted loss that jointly captures electricity cost and a penalty for workload shedding, thereby evaluating scheduling decision quality in terms of both economy and system security. 
   Case studies demonstrate that the proposed approach yields lower operational costs and higher constraint satisfaction compared to the conventional two-stage baseline.
\end{enumerate*}

\section{Model Formulation}
\label{sec:model_formulation}
This section introduces the mathematical models for the power distribution network and the AI data center, and formulates a grid-coordinated spatio-temporal workload scheduling optimization problem.

\subsection{Power Distribution Network Modeling}
\label{sec:power_grid_model}
{
We consider a multi-period scheduling horizon denoted by the set $\mathcal{T} = \{1, \dots, T\}$, where each time step $t \in \mathcal{T}$ represents a discrete interval with a resolution of $\Delta t = 1$ hour.
}
We model the power distribution network using the linearized DistFlow model \cite{Farivar2013Branch}. In this framework, the AI datacenters are treated as specialized, controllable loads. Each datacenter is connected to a specific bus $i$ within the distribution network, denoted as $i \in \mathcal{N}_{\DC}$. The core of the grid model is to maintain power balance at each bus. This is expressed through the active and reactive power balance constraints \eqref{eq:active_power_flow_constr} and \eqref{eq:reactive_power_flow_constr}:
{\small
\begin{equation}
\setlength\abovedisplayskip{3pt}
\setlength\belowdisplayskip{3pt}
\sum_{g \in \Omega_i^G} P^{g}_{i,t}-
\sum_{j \in \Omega_i^{\text{out}}} P^{l}_{ij,t}+
\sum_{j \in \Omega_i^{\text{in}}} P^{l}_{ji,t}  =
\begin{cases}
P^{\DC}_{i,t}, \text{if } i \in \mathcal{N}_{\DC}, \\
P^{d}_{i,t}, \text{otherwise},
\end{cases}
\label{eq:active_power_flow_constr}
\end{equation}
}%
{\small
\begin{equation}
\sum_{g \in \Omega_i^G} Q^{g}_{i,t}-
\sum_{j \in \Omega_i^{\text{out}}} Q^{l}_{ij,t}+
\sum_{j \in \Omega_i^{\text{in}}} Q^{l}_{ji,t} 
=
\begin{cases}
0, \text{if } i \in \mathcal{N}_{\DC},\\
Q^{d}_{i,t}, \text{otherwise},
\end{cases}
\label{eq:reactive_power_flow_constr}
\end{equation}
}%
where $P^{g}_{i,t}$ and $Q^{g}_{i,t}$ are the active and reactive power generation, while $P^{l}_{ij,t}$ and $Q^{l}_{ij,t}$ are the line power flows. 
{
The terms $P_{i,t}^{d}$ and $Q_{i,t}^{d}$ represent the conventional active and reactive background loads at bus $i$, respectively.
}
~\eqref{eq:active_power_flow_constr} and ~\eqref{eq:reactive_power_flow_constr} couple the total power consumption $P^{\DC}_{i,t}$ of the AI datacenter into the power balance equation  at bus $i$.

The model is completed by the voltage drop constraint \eqref{eq:voltage_drop_constr} and physical limit constraints \eqref{eq:slack_bus_voltage_constr}-\eqref{eq:p_gen_bound_constr}, which enforce limits on bus voltages and power generation:
{\small
\label{eq:slack_bus_voltage_constr}
\begin{align}
&V_{i,t} - V_{j,t} =
2(r_{ij}\cdot P_{ij,t}^{l}
+ x_{ij}\cdot Q_{ij,t}^{l}),
\label{eq:voltage_drop_constr}\\
&V_{0,t} = v_{r}^{2}, \quad \underline{v}_{i}^{2} \leq V_{i,t} \leq \overline{v}_{i}^{2},\label{eq:slack_bus_voltage_constr}\\
\label{eq:p_gen_bound_constr}
&\underline{P}^{g} \leq P^{g}_{i,t} \leq \overline{P}^{g}, \quad
\underline{Q}^{g} \leq Q^{g}_{i,t} \leq \overline{Q}^{g}.
\end{align}

\subsection{AI Datacenter Power Consumption Modeling}
\label{sec:dc_power_model}
The power consumption of an AI datacenter is primarily driven by its IT equipment load, $P^{\IT}_{i,t}$, which represents the power consumed by servers and GPUs processing computational tasks. This IT load has a distinct operational range, physically bounded by the cluster's idle power $P^{\idle}_i$ and its maximum power capacity $P^{\full}_i$:
{\small
\begin{equation}
\setlength\abovedisplayskip{3pt}
\setlength\belowdisplayskip{3pt}
P^{\idle}_i \le P^{\IT}_{i,t} \le P^{\full}_i, \quad \forall  i \in \mathcal{N}_{\DC}.
\label{eq:dc_power_upper_constr}
\end{equation}
}

However, $P^{IT}_{i,t}$ only accounts for the computational hardware. The {total} electricity demand drawn from the power grid, $P^{\DC}_{i,t}$, must also account for auxiliary systems such as cooling, lighting, and power distribution units. We utilize the standard Power Usage Effectiveness (PUE) metric $R_{\PUE}$, to model this relationship and map the IT load to the total cluster demand:
{\small
\begin{equation}
\setlength\abovedisplayskip{3pt}
\setlength\belowdisplayskip{3pt}
P^{\DC}_{i,t} = R_{\PUE} \cdot P^{\IT}_{i,t}, \quad \forall i \in \mathcal{N}_{\DC}.
\label{eq:p_dc_by_p_IT_constr}
\end{equation}
}This variable, $P^{\DC}_{i,t}$, represents the total load that the datacenter imposes on the power grid, as incorporated in the power balance equation \eqref{eq:active_power_flow_constr}.

\subsection{AI Datacenter Spatio-temporal Workload Scheduling}
\label{sec:downstream}
Many tasks in AI datacenters, such as deep learning jobs, are delay-tolerant, which endows their workloads with inherent spatio-temporal flexibility. We model this scheduling capability via a transfer matrix $M$. An element $M_{j,i,s,t}$ represents the percentage of workload that arrived at datacenter $j$ at time $s$ and is scheduled for execution at datacenter $i$ at time $t$. This decision matrix must adhere to operational constraints:
{\small
\begin{subequations}
\label{eqs:power_transfer}
\setlength\abovedisplayskip{3pt}
\setlength\belowdisplayskip{3pt}
\begin{align}
\label{eq:transfer_bound_constr}
&M_{j,i,s,t} \in [0,1],  \quad \forall i,j \in \mathcal{N}_{\DC}, \; s,t \in \mathcal{T},\\
\label{eq:transfer_bound_sum_constr}
&\sum_{i \in \mathcal{N}_{\DC}} \sum_{t \in \mathcal{T}} M_{j,i,s,t} = 1, \quad \forall j \in \mathcal{N}_{\DC}, \; s \in \mathcal{T},\\
\label{eq:transfer_time_constr}
&M_{j,i,s,t} = 0, \quad \text{if } \; t < s,
\end{align}
\end{subequations}}\eqref{eq:transfer_bound_sum_constr} enforces load conservation, ensuring all submitted tasks are dispatched, while \eqref{eq:transfer_time_constr} ensures temporal feasibility. Given a predicted workload $\hat{w}_{j,s}$ from an upstream model, the scheduling decision $M$ determines the realized IT power consumption:
{\small
\begin{equation}
\setlength\abovedisplayskip{3pt}
\setlength\belowdisplayskip{3pt}
P^{\IT}_{i,t} = \sum_{j \in \mathcal{N}_{\DC}} \sum_{s \in \mathcal{T}}
\hat{w}_{j,s} \cdot M_{j,i,s,t} + P^{\idle}_i.
\label{eq:IT_power_after_transfer_constr}
\end{equation}}%
By combining the models from Sections \ref{sec:power_grid_model}, \ref{sec:dc_power_model}, and \ref{sec:downstream}, we formulate the downstream grid-coordinated workload scheduling problem. The objective is to minimize the total electricity generation cost \eqref{eq:opt_obj}, subject to the complete set of system constraints:
{\small
\begin{subequations}
\setlength\abovedisplayskip{3pt}
\setlength\belowdisplayskip{3pt}
\begin{align}
 &\min \limits_{\begin{small} \begin{subarray}{c} M, P^{{\IT}}, P^{\DC}\\P^{g}, Q^{g}, P^{l}, Q^{l}, V\end{subarray} \end{small}} \sum_{t \in \mathcal{T}} \sum_{i \in \Omega^{g}} \pi _{i,t} \cdot P^{g}_{i,t}
\label{eq:opt_obj}
\\
 &\qquad \text{s.t.}\quad \eqref{eq:active_power_flow_constr}-\eqref{eq:IT_power_after_transfer_constr},\nonumber
\end{align}
\label{eq:downstream_problem}\end{subequations}}where $\pi_{i,t}$ is the electricity price. 
The constraints are grouped as follows:~\eqref{eq:active_power_flow_constr}-\eqref{eq:p_gen_bound_constr} represent the {power grid constraints}, ensuring power flow balance and voltage limits.~\eqref{eq:dc_power_upper_constr}-\eqref{eq:p_dc_by_p_IT_constr} are the {datacenter power constraints}, linking IT load to total grid consumption.~\eqref{eqs:power_transfer}-\eqref{eq:IT_power_after_transfer_constr} are the {workload transfer constraints}, which define the spatio-temporal scheduling logic and its impact on IT load.

\begin{figure*}[t] 
\centering 
\includegraphics[width=0.8\linewidth]{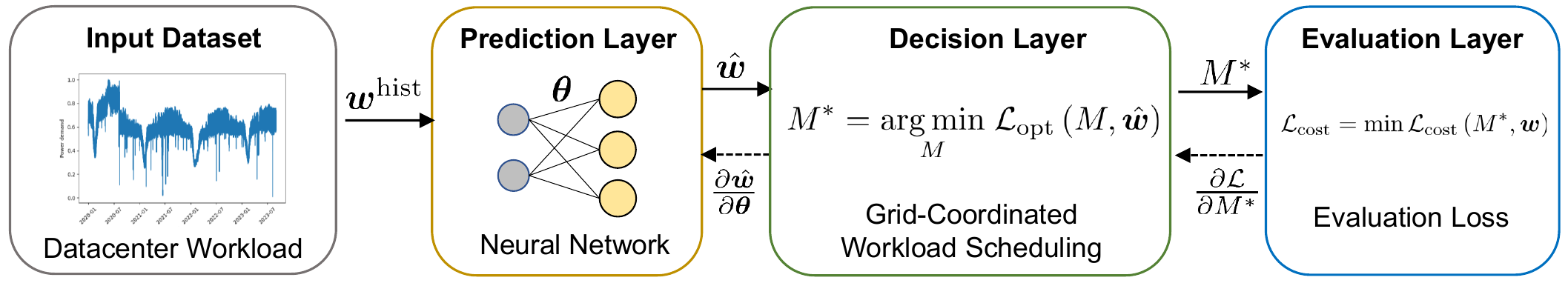} 
\caption{End-to-end Predict-then-Schedule Framework} 
\label{fig:pto_datacenter_plot} 
\vspace{-0.7cm}
\end{figure*}

\section{Methodology}
\label{sec:methodology}
This section details the proposed PTS learning framework, which integrates workload prediction, scheduling optimization, and decision evaluation into a unified closed-loop, end-to-end training pipeline.

\subsection{The End-to-End Predict-then-Schedule Framework}
\label{sec:upstream} 
In conventional two-stage approaches, prediction and optimization are treated as two separate, sequential procedures. First, an upstream prediction model is trained independently to minimize a statistical loss, such as Mean Squared Error (MSE) \eqref{eq:mse_loss}. Its predictions $\hat{w}$, are then treated as fixed, deterministic inputs for the downstream scheduling problem. This approach is limited by its failure to account for the impact of prediction errors on the final decision quality.
{\small\begin{equation}
\setlength\abovedisplayskip{3pt}
\setlength\belowdisplayskip{3pt}
\label{eq:mse_loss}
\mathcal{L}_{\MSE}(\theta) = \frac{1}{N} \sum_{\{(w^{\hist}_i, w_i)\} \in \mathcal{D}} \| \hat{w}_i - w_i\|^2, \; \hat{w}_i = f(w^{\hist}_i; \theta),
\end{equation}}%
where $w^{\hist}_i$ is the historical workload used as input features, $w_i$ is the corresponding ground-truth future workload, and $(w^{\hist}_i, w_i)$ represents a training sample drawn from the dataset $\mathcal{D}$.
To overcome this limitation, we propose an end-to-end, decision-oriented PTS framework.  We integrate the downstream grid-coordinated scheduling problem \eqref{eq:downstream_problem} directly into the upstream model's training pipeline by treating it as a differentiable decision layer. Rather than treating $\hat{w}$ as a constant, the framework views it as a differentiable parameter. This allows decision-based gradients from the final evaluation in Section \ref{sec:redispatch_loss} to backpropagate through the optimization layer via implicit differentiation in Section \ref{subsec:implicit-diff}. Consequently, the prediction model learns to produce decision-focused forecasts specifically optimized for high-quality, low-cost downstream scheduling.

\subsection{Decision Evaluation Model with Over-shifted Loss}
\label{sec:redispatch_loss} 
The Decision Evaluation Layer quantifies the true operational cost and ensures the differentiability of the learning pipeline.
It evaluates the scheduling decision $M^*$, generated based on the prediction $\hat{w}$, against the ground-truth workload $w$.
In the evaluation phase, the realized IT load $P^{\IT*}_{i,t}$ upon $w$ and $M^{*}$ is defined as:
{\small
\begin{equation}
\setlength\abovedisplayskip{3pt}
\setlength\belowdisplayskip{3pt}
\label{eq:p_task_realized}
P^{\IT*}_{i,t}
= \sum_{j \in \mathcal{N}_{\DC}} \sum_{s \in \mathcal{T}}
w_{j,s} \cdot M^*_{j,i,s,t} + P^{\idle}_i.
\end{equation}}%
A fundamental challenge arises from the discrepancy between $\hat{w}$ and $w$.
If $M^*$ under-predicts the load and $P^{\IT*}_{i,t}$ exceeds the capacity $P^{\full}_i$, a strict optimization problem becomes infeasible, causing the gradient flow to break.
\par
To guarantee the evaluation problem remains solvable, we relax the hard capacity constraint by introducing a load shedding variable $P^{\shed}_{i,t}$ and propose the Over-shifted Loss $\mathcal{L}_{\cost}$, which heavily penalizes any shedding.
The evaluation is formulated as:
{\small
\begin{subequations}
\setlength\abovedisplayskip{3pt}
\setlength\belowdisplayskip{3pt}
\begin{align}
\mathcal{L}_{\cost} = &\min\limits_{\Xi} \sum_{t \in \mathcal{T}}
\left(
\sum_{i \in \Omega^{g}} \pi _{i,t} P^{g}_{i,t} +
\sum_{i \in \mathcal{N}_{\DC}} \gamma \left(P^{\shed}_{i,t}\right)^2
\right)
\label{eq:eval_obj}\\
\text{s.t.}\quad & \eqref{eq:active_power_flow_constr} - \eqref{eq:p_dc_by_p_IT_constr} \nonumber\\
& P^{\IT}_{i,t} = P^{\IT*}_{i,t} - P^{\shed}_{i,t},\quad P^{\shed}_{i,t} \ge 0. \label{eq:shed_nonneg}
\end{align}
\label{eq:decision_evaluation}
\end{subequations}}
Here $\Xi$ denotes the set of optimization variables. 
The coefficient $\gamma > 0$ is a penalty factor that governs the trade-off between economic optimality and constraint satisfaction. 
A sufficiently large $\gamma$ prioritizes the elimination of load shedding, incentivizing the model to produce predictions that inherently hedge against grid violations.
Constraints \eqref{eq:shed_nonneg} replace the hard capacity constraint with a soft relaxation to maintain gradient continuity during backpropagation. 
Minimizing \eqref{eq:eval_obj} explicitly teaches the upstream model to avoid forecasts that lead to unsafe schedules.

\subsection{End-to-End Training via Implicit Differentiation}
\label{subsec:implicit-diff} 
The key technical challenge in realizing the end-to-end framework from Section \ref{sec:methodology} is enabling the backpropagation of gradients from the $\mathcal{L}_{\cost}$ defined in Section \ref{sec:redispatch_loss} {through} the decision layer to the parameters $\theta$ of the upstream predictor.
We achieve this using the implicit differentiation. This technique is applicable because our downstream decision problems are convex programs with linear equality and inequality constraints. Drawing on foundational work such as \cite{amos2017optnet,agrawal2019differentiable}, the process works as follows:
\subsubsection{Compact form}
For convenience, we rewrite the decision layer in a generic form. Let $y$ stack all primal decision variables, e.g., $\mathrm{vec}(M)$, network flows, voltages, generations, and let $\phi$ collect problem parameters, e.g., the predicted load $\hat w(\theta)$ and other exogenous coefficients. Denote the convex objective by $J(y,\phi)$, the linear equalities by $h(y,\phi)=0$, and the linear inequalities by $g(y,\phi)\le 0$. The problem is
{\small
\begin{equation}
\setlength\abovedisplayskip{3pt}
\setlength\belowdisplayskip{3pt}
\min_{y} J(y,\phi)\quad
\text{s.t.} \quad h(y,\phi)=0,\quad g(y,\phi)\leq 0 .
\label{eq:compactQP}
\end{equation}
}

\subsubsection{Lagrangian and KKT system}
Let $\lambda$ and $\mu\!\ge\!0$ be the dual variables for $f(\cdot)=0$ and $g(\cdot)\le0$, respectively. The Lagrangian is
{\small
\setlength\abovedisplayskip{3pt}
\setlength\belowdisplayskip{3pt}
\begin{equation}
\mathcal{L}(y,\lambda,\mu;\phi)
= J(y,\phi)+\lambda^{\top}f(y,\phi)+\mu^{\top}g(y,\phi).
\label{eq:Lagrangian}
\end{equation}
}Because \eqref{eq:compactQP} is convex and satisfies standard regularity, strong duality holds and the KKT conditions are necessary and sufficient for optimality. The KKT system can be expressed as the vector equation
{\small
\begin{equation}
\setlength\abovedisplayskip{3pt}
\setlength\belowdisplayskip{3pt}
\mathcal{F}(y^{*},\phi)=
\begin{bmatrix}
\nabla_{y} \mathcal{L}(y^{*},\lambda^{*},\mu^{*};\phi)\\
h(y^{*},\phi)\\
D(\mu^{*})\,g(y^{*},\phi)
\end{bmatrix}
= 0,
\; \mu^{*} \geq 0, \; g(y^{*},\phi) \leq 0.
\label{eq:KKT}
\end{equation}
}
where $D(\mu)$ denotes the diagonal matrix formed by $\mu$.
\subsubsection{Implicit differentiation}
By the implicit function theorem, the optimizer $y^{*}$ is a differentiable function of $\phi$ and its sensitivity is obtained by differentiating \eqref{eq:KKT}:
{\small
\begin{equation}
\setlength\abovedisplayskip{3pt}
\setlength\belowdisplayskip{3pt}
\frac{\partial \mathcal{F}}{\partial \phi}
+
\frac{\partial \mathcal{F}}{\partial y}
\frac{\mathrm{d}y^{*}}{\mathrm{d}\phi}
=0
\quad\Rightarrow\quad
\frac{\mathrm{d}y^{*}}{\mathrm{d}\phi}
=
-\Big[\tfrac{\partial \mathcal{F}}{\partial y}\Big]^{-1}
\Big[\tfrac{\partial \mathcal{F}}{\partial \phi}\Big].
\label{eq:IFT}
\end{equation}
}Here $\tfrac{\partial \mathcal{F}}{\partial y}$ is the nonsingular KKT matrix composed of the Hessian blocks of \eqref{eq:Lagrangian} and the constraint Jacobians under LICQ and strict complementarity. Instead of forming an explicit inverse, we solve the linear system defined by \eqref{eq:IFT} to obtain Jacobian–vector products efficiently.

\subsubsection{Backpropagation to the predictor}
Since the parameters $\phi$ depend on the predictor via $\phi=\hat w(\theta)$, the gradient of the over-shifted loss $\mathcal{L}_{\cost}$  in \eqref{eq:eval_obj} with respect to the prediction network weights $\theta$ follows from the chain rule:
{\small
\begin{equation}
\setlength\abovedisplayskip{3pt}
\setlength\belowdisplayskip{3pt}
\frac{\mathrm{d}\mathcal{L}_{\cost}}{\mathrm{d}\theta}
=
\frac{\mathrm{d}\mathcal{L}_{\cost}}{\mathrm{d} y^{{*}}}
\frac{\mathrm{d} y^{{*}}}{\mathrm{d} \phi}
\frac{\mathrm{d} \phi }{\mathrm{d} \theta}.
\label{eq:chain}
\end{equation}
}\eqref{eq:KKT}–\eqref{eq:chain} enable end-to-end training of the predict then schedule pipeline. The downstream optimizer remains an exact convex program, while its solution is fully differentiable with respect to the upstream forecast parameters.

\subsection{Complete Training Framework}
\label{sec:complete_framework}
In summary, our proposed PTS architecture adopts the three-layer, end-to-end framework illustrated in Fig.~\ref{fig:pto_datacenter_plot}. This framework connects workload prediction, decision-making, and decision evaluation in a closed loop. The layers are:
\begin{itemize}
    \item {Prediction Layer (Section~\ref{sec:upstream}):} An upstream neural network $f(w^{\hist};\theta)$, implemented as an LSTM, maps system features $w^{\hist}$ to a workload prediction $\hat{w}$.
    
    \item {Decision Layer (Section~\ref{sec:downstream}):} A differentiable convex optimization layer that takes the prediction $\hat{w}$ as a parameter, solves the grid-coordinated scheduling problem \eqref{eq:downstream_problem}, and outputs the optimal decision $M^*$.
    
    \item {Decision Evaluation Layer (Section~\ref{sec:redispatch_loss}):} This layer assesses the {true} quality of the decision $M^*$ by testing it against the ground-truth workload $w$. It solves the decision evaluation problem \eqref{eq:decision_evaluation} to compute the final {over-shifted loss $\mathcal{L}_{\cost}$}, which quantifies both economic cost and penalties for any safety-violating infeasibilities.
\end{itemize}

During training, this end-to-end pipeline enables the $\mathcal{L}_{\cost}$ to be backpropagated from the evaluation layer, {through} the differentiable decision layer via Section \ref{subsec:implicit-diff}, and {into} the prediction layer to update $\theta$. This feedback loop trains a predictor that is inherently aware of the downstream task. It learns to produce forecasts that proactively avoid high-cost, operationally infeasible regions of the solution space, thereby directly minimizing the true operational objectives of cost and safety.

\section{Simulation}

To validate the proposed end-to-end PTS framework, we evaluate its performance against the conventional two-stage method in terms of operational cost, prediction accuracy, and constraint satisfaction.

\subsection{Experimental Setup}
\subsubsection{System Configuration}
Simulations are performed on a modified IEEE 33-bus distribution test system. One AI data center is integrated at bus $24$. 
The parameters of the AI data center include an idle power consumption of $0.00$ units and a full power consumption capacity of $0.15$ units. The scheduling horizon is $24$ hours with hourly discretization. The hourly electricity price profile, which is shown in Fig.~\ref{fig:transfer_comparison}, is generated through a transformation of a sinusoidal function, with values ranging between 0.5 and 1, reflecting the diurnal fluctuations in electricity prices.

\subsubsection{Data Preprocessing and Model Parameters}
The workload dataset used in this experiment is synthesized from real-world power demand forecasting data and has been normalized to [0,1], which is shown in Fig.~\ref{fig:1108load}.
It consists of hourly power load values that represent the electrical consumption. 
\begin{figure}[!t]
    \centering
    \includegraphics[width=0.7 \linewidth]{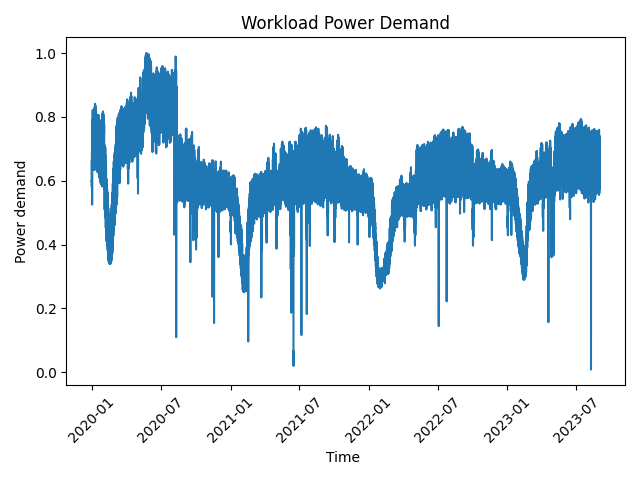}
    \vspace{-0.4cm}
    \caption{Our Workload Dataset}
    \label{fig:1108load}
\end{figure}
{
This dataset spans 3.5 years of hourly data, totaling approximately 30,660 data points. This duration was selected to capture diverse seasonal and diurnal load patterns for robust generalization, and is partitioned into $80\%$ for training and $20\%$ for testing.
}

\subsubsection{Models for Comparison}
We benchmark our proposed approach against the traditional two-stage baseline, evaluating performance across various metrics:
\begin{itemize}
   \item {Two-Stage: The traditional baseline framework, where the predictive model is trained solely to minimize MSE in \eqref{eq:mse_loss}.}

    \item {Two-Stage (Reserve): This method sets a new operational limit at 80\% of the rated capacity while reserving the remaining 20\% as a safety margin.}

    \item {PTS}:
    Our proposed framework, where the predictor is trained to minimize the over-shifted loss~\eqref{eq:eval_obj} that balances operational cost and power system security.
\end{itemize}

\subsubsection{Implementation and Training Details}
The prediction model is a LSTM neural network implemented in PyTorch. The downstream optimization is formulated in CVXPY and embedded as a differentiable layer. Experiments are conducted on an HPC node equipped with an Intel Core i5-13600K CPU and one NVIDIA RTX-3060 GPU. Models are trained for $20$ epochs with learning rate of $1 \times 10^{-3}$ and batch size of $30$.

\subsection{Results and Analysis}
We evaluate the two training strategies on the test dataset, comparing operational cost, prediction accuracy, and the quality of the resulting scheduling decisions.

\begin{figure}[t]    
    \centering
    
    \begin{subfigure}[t]{0.35 \textwidth}
        
        \centering
        \includegraphics[width=  \textwidth]{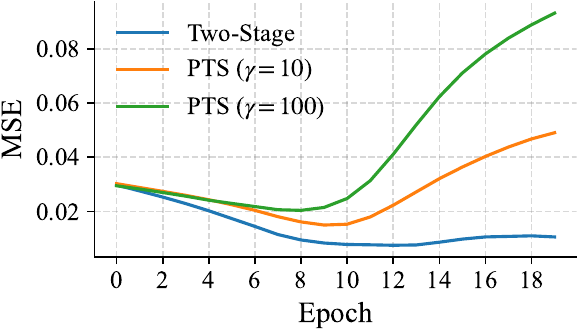}
        \caption{\footnotesize The mean squared
error between predicted and ground-truth workload among three models.  }
        \label{fig:mse_comparison}
    \end{subfigure}
    \begin{subfigure}[t]{0.35 \textwidth}
        \centering
        \includegraphics[width=  \textwidth]{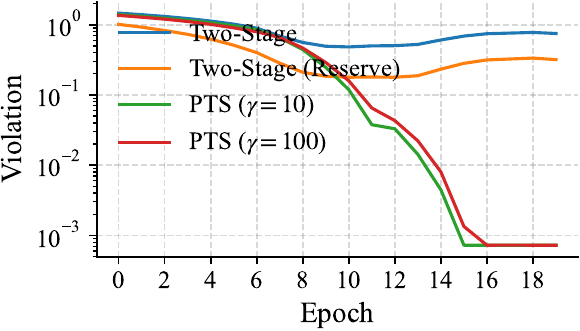}
        \caption{\footnotesize The magnitude of constraint violations among four methods.   }
        \label{fig:viol_comparison}
    \end{subfigure}
    \begin{subfigure}[t]{0.35 \textwidth}
        \centering
        \includegraphics[width= \textwidth]{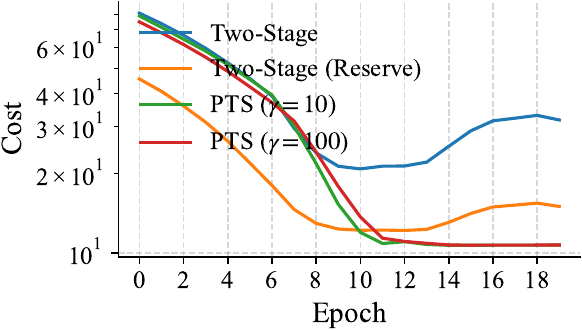}
        \caption{\footnotesize The cost comparison among four methods.  }
        \label{fig:cost_comparison}
    \end{subfigure}
    \begin{subfigure}[t]{0.35 \textwidth}
        \centering
        \includegraphics[width=  \textwidth]{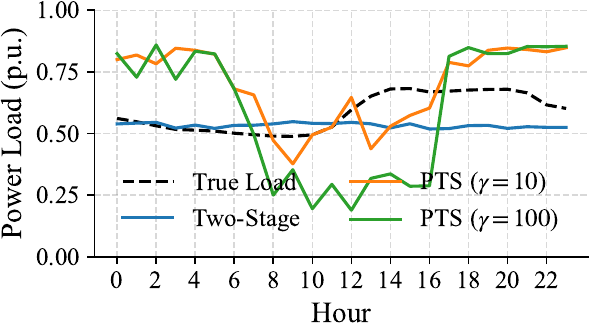}
        \caption{\footnotesize The prediction comparison between true load and predicted load from three models. }
        \label{fig:load_pred_comparison}
    \end{subfigure}
    
    \caption{Performance Comparison of Different Methods}    
    \label{fig:comparison_all}
    \vspace{-0.5cm}
\end{figure}

\begin{table}[]
\centering
\caption{Comparison of metrics among different training methods.}
\label{tab:metrics_comparison} 
\begin{tabular}{@{}llll@{}}
\toprule
Method & MSE Loss & Cost Loss & Violation   \\ \midrule
Two-Stage    & 0.0106   & 31.8381   & 0.7518    \\
Two-Stage (Reserve)  &0.0106  & 14.9977 & 0.3190\\
PTS ($\gamma$=10)   & 0.0490   & 10.7189  & 0.0007      \\ 
PTS ($\gamma$=100)   & 0.0931   & 10.7205  & 0.0007      \\ \bottomrule
\end{tabular}
\vspace{-0.5cm}
\end{table}

\subsubsection{Operational Cost and Accuracy Trade-off}


The comparative performance during training is evaluated across three dimensions. Figure \ref{fig:viol_comparison} illustrates the magnitude of constraint violations, Figure \ref{fig:mse_comparison} displays the mean squared error between predicted and ground-truth workloads, and Figure \ref{fig:cost_comparison} presents the total operational cost. Table \ref{tab:metrics_comparison} further summarizes the final quantitative results.

The results reveal a distinct trade-off between statistical fidelity and decision quality. 
While the Two-Stage model achieves the lowest MSE, it suffers from significant constraint violations and high operational costs.
The Reserve baseline improves system safety relative to the standard Two-Stage approach but results in suboptimal costs due to its inflexible capacity limits. 
In contrast, the PTS models prioritize the more critical downstream decision objective over conventional statistical accuracy. 
The prediction MSE increases in the later stages of training and thus successfully reduces the violation metric by nearly an order of magnitude to a stable and near-zero level. 
This performance demonstrates that internalizing downstream constraints and cost objectives enables the model to generate safe, feasible, and significantly more cost-effective decisions.

The sensitivity analysis regarding the penalty coefficient $\gamma$ further demonstrates the structural robustness of the proposed framework. 
As illustrated in Fig.~\ref{fig:viol_comparison}, both $\gamma=10$ and $\gamma=100$ successfully reduce violations to a comparable stable floor. Although a larger $\gamma$ of 100 imposes a stricter penalty that leads to higher initial operational costs in Fig.~\ref{fig:cost_comparison}, both configurations eventually converge to similar cost levels. This outcome indicates that the framework is insensitive to the specific value of $\gamma$ provided the coefficient is sufficiently large to maintain system constraints.


\subsubsection{Analysis of Prediction Behavior}

As shown in Fig.~\ref{fig:load_pred_comparison}, the predictive behaviors differ significantly. 
The Two-Stage model adheres closely to the true load, displaying high statistical fidelity. 
In contrast, the PTS models produce decision-oriented predictions with intentional deviations, for example, over-forecasting during hours 0-6, 17-23 and under-forecasting during hours 11-16. 
This sacrifice in statistical accuracy is not a flaw, but rather a learned strategy, as the {PTS} models recognize this distorted prediction is essential for guiding the downstream optimization process away from infeasible or high-cost decisions.

\begin{figure}[!t]
    \centering
    \includegraphics[width=0.98\linewidth]{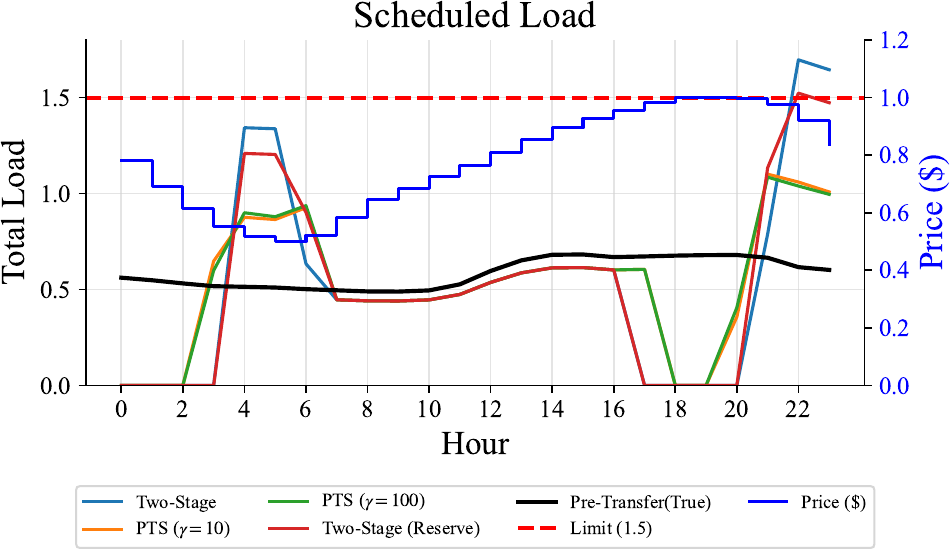}
    \caption{Comparison of task transfer results among four methods.}
    \label{fig:transfer_comparison}
\end{figure}

\subsubsection{Impact on Scheduling Decisions}

The difference in predictive behavior directly impacts the feasibility of the downstream scheduling decisions, as shown in Fig.~\ref{fig:transfer_comparison}. The Two-Stage models, based on its statistically accurate forecast, attempts to exploit low prices during hours 4-6, 22-23, but this results in severe violations of the capacity limit. The Reserve method avoids extreme violations but shows a minor violation. Conversely, the PTS framework demonstrates its superiority. By leveraging its awareness of the downstream optimization, it accurately identifies the capacity limit as a critical bottleneck. Its distorted prediction was specifically designed to avoid this. The final schedule intelligently smooths the workload scheduling, both exploiting the low prices and, crucially, never breaching the operational limit.

\begin{figure}[!t]
    \hspace{0.4cm}
\includegraphics[width=0.98\linewidth]{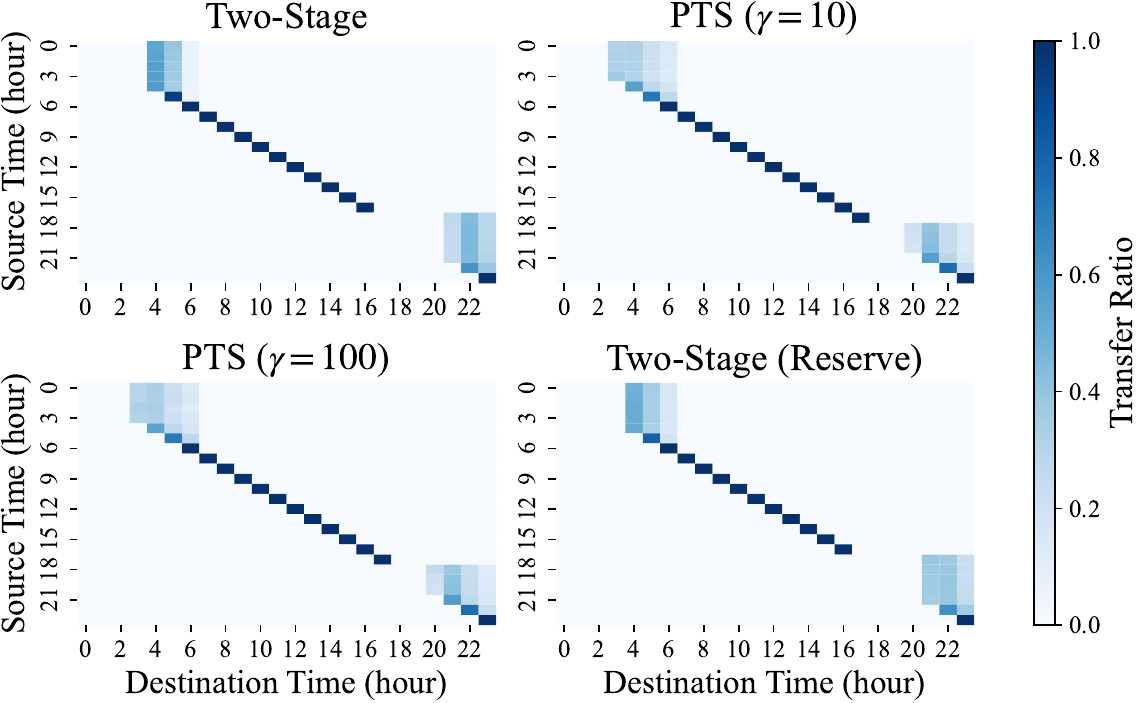}
    \vspace{-0.4cm}
    \caption{Heatmaps of learned transfer matrices $M_{d,c,s,t}$ among four methods.}
    \label{fig:transfer_heatmap}
\end{figure}

\subsubsection{Analysis of Learned Transfer Policies}

Fig.~\ref{fig:transfer_heatmap} visualizes the learned transfer matrices $M$. While both models learn to transfer workload to low-cost periods, the policy learned by the PTS model holds relatively conservative transfer strategies. 
For instance, to prevent constraint violations, the PTS will not shift the workload from 17:00 to later time slots, even if this could improve overall resource utilization.

\section{Conclusion}
This paper introduces an end-to-end PTS framework for AI datacenter workload scheduling coordinated with power distribution networks. By integrating the downstream convex scheduling optimization as a differentiable layer using implicit differentiation, our framework aligns workload prediction with final operational objectives. The proposed workload over-shifted loss ensures that the model is trained to balance operational cost with power system security. Experimental results demonstrate that our framework significantly reduces operational cost and improves constraint satisfaction compared to the conventional two-stage baseline.

\bibliographystyle{IEEEtran}
\bibliography{references}

\end{document}